\documentclass[12pt]{article}
\usepackage{latexsym,graphicx,amssymb,color,amsmath}
\usepackage{epsfig}
\usepackage{lscape}
\date{}
\graphicspath{}
\DeclareGraphicsExtensions{.eps}

\newcommand{\N}{\mathbb{N}}

\newcommand{\be}{\begin{equation}}
\newcommand{\ee}{\end{equation}}
\newcommand{\barray}[1][rcl]{\be\begin{array}{#1}}
\newcommand{\barrayl}[2][rcl]{\be\label{#2} \begin{array}{#1}}
\newcommand{\earray}{\end{array}\ee}

 {\hfill{$\Box$}\bigskip\par}%

\begin{document}

\vspace*{-2cm}
\begin{flushright}
\end{flushright}

\vspace{0.3cm}

\begin{center}
{\Large {\bf A Note on Genome Organisation in RNA Viruses with Icosahedral Symmetry}}\\ 
\vspace{1cm} {\large \bf N.\ Jonoska\,${}^{}$\footnote{\noindent E-mail: 
{\tt jonoska@math.usf.edu}} and
R.\ Twarock\,${}^{}$\footnote{\noindent E-mail: 
{\tt rt507@york.ac.uk}}}\\
\vspace{0.3cm} {${}^1$}\em Department of Mathematics \\ University of South Florida, Tampa, FL 33620, USA\\

\vspace{0.3cm} {${}^2$} Departments of Mathematics and Biology \\ University of York, York YO10 5DD, U.K.\\ 
\end{center}

\begin{abstract}
\noindent 
The structural organisation of the viral genome within its protein container, called the viral capsid, is an 
important aspect of virus architecture. Many single-stranded (ss) RNA viruses organise a significant part 
of their genome in a dodecahedral cage as a RNA duplex structure that mirrors the symmetry of the 
capsid. Bruinsma and Rudnick have suggested a model for the structural organisation of the RNA in 
these cages. It is the purpose of this paper to further develop their approach based on results from the 
areas of graph theory and DNA network engineering. We start by suggesting a scenario for pariacoto 
virus, a representative of this class of viruses, that is energetically more favorable than those derived 
previously. We then show that it is a representative of a whole family of cage structures that abide to the 
same construction principle, and then derive the energetically optimal configuration for a second family 
of cage structures along similar lines. Finally, we give reasons for the conjecture that these two families 
are more likely to occur in nature than other scenarios.   
\end{abstract}

One of the important open problems in virology is the uncovering of the structural organisation of the 
viral genome within the protein containers or viral capsids that encapsulate and hence provide 
protection for the viral genome. Experiments have shown that the packaging structures of different types 
of viruses are distinct, ranging from spool-like organisations in some dsDNA viruses \cite{Odijk} to an 
organisation (of a portion of the viral genome) in terms of dodecagonal cages in certain RNA viruses. 
We focus here on the latter, and in particular consider pariacoto virus, a single stranded RNA virus  in 
the family of Nodaviridae, as a representative of viruses in this class. Tang {\it et al.} have shown 
\cite{Tang} that about 35$\%$ of the genomic viral RNA of this virus are organised in a dodecahedral 
cage as a RNA duplex structure. A theoretical model for the structural organisation of the RNA in this 
cage structure has been pioneered in \cite{Bruinsma}. 
It models the location of the RNA molecule as a directed path on a dodecahedral graph that visits each 
edge precisely twice in opposite direction, hence generating the double helices observed in \cite{Tang} 
with the two strands oriented in the standard way. The theory relies on a set of vertex configuration that 
describe potential ways of organising the RNA at the three-coordinated vertices of the dodecahedral 
cage. In this note, we extend this set based on results from the enineering of DNA networks
(construction of arbitrary DNA graph structures \cite{JACS}), where another type of ``junction'' has been shown to occur. We use this extended set of vertex configurations, together with the results on double strand numbers on thickened graphs in \cite{Jonoska} to derive a model for the RNA organisation in pariacoto virus that is energetically more favourable than those derived previously. 

\begin{figure}[ht]
\begin{center}
\includegraphics[width=15cm]{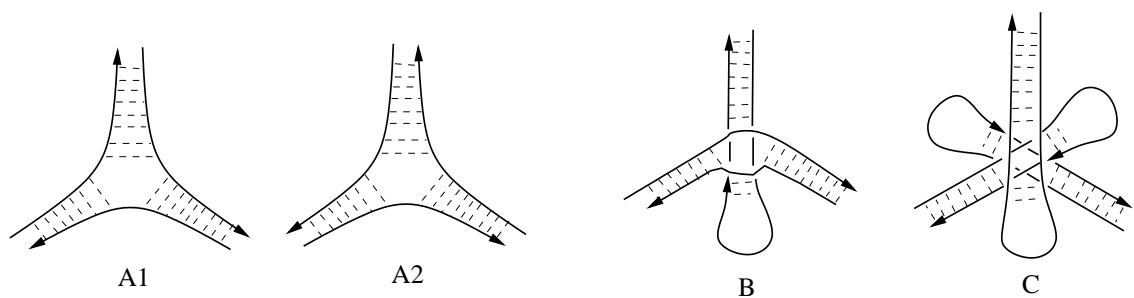}
\end{center}
\caption{The vertex configurations suggested in \cite{Bruinsma}.}
\label{Fig1}
\end{figure}

Our starting point are the vertex configurations shown in Fig. \ref{Fig1}, that have been adapted from 
\cite{Bruinsma}, and will be called branch point (A1 and A2), bubble/hairpin (B) and triple hairpin (C), 
respectively, following the terminology introduced in this reference. 
The respective energies of these configurations have been discussed in detail in \cite{Bruinsma}, and 
we therefore only point out here that the configurations A1 and A2 are energetically more favourable 
than B, which in turn is more favourable than C, with  a   potential energy difference of the order of magnitude of 
$10^2 k_BT$ between these options. Moreover, the authors have concluded that the energy difference 
between competing dodecahedral arrangements are determined by the choice of the vertex types. It is 
therefore  important to consider a suitable set of vertex configurations. 

Studies of DNA networks have shown that there is a further possible 
organization at the vertices \cite{JACS}. 
\begin{figure}[ht]
\begin{center}
\includegraphics[width=6cm,keepaspectratio]{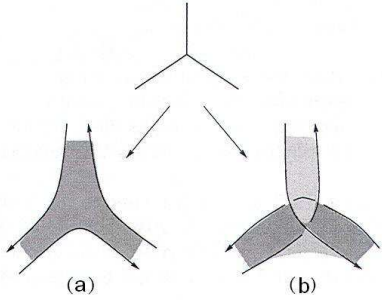}
\end{center}
\caption{The new vertex configuration (b) shown in comparison with a branch point of type (A1) or (A2) in (a). Figure adapted from \cite{Jonoska}.}
\label{Fig2}
\end{figure}
It is illustrated schematically in Fig. \ref{Fig2}(b) in comparison with the vertex configuration of type A. 
We believe that  it is   energetically more favourable than the bubble/hairpin or tripple hairpin vertex types in Fig. \ref{Fig1}.  In fact its energy  should be very similar to that of vertex type A, as it has a very similar structure   
which we demonstarate in a close-up view in Fig.~\ref{Ajunction}.
\begin{figure}[ht]
\begin{center}
\includegraphics[width=10cm,keepaspectratio]{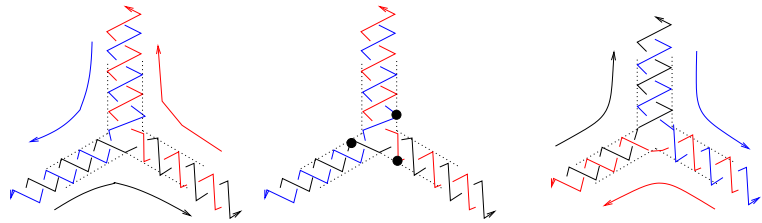}
\end{center}
\caption{The difference between vertex type A (left) and the new type in Fig.~\ref{Fig2} (right) demonstrated in a close-up. Differences in length of the RNA sequence of only about 5 nucleotides at the locations indicated by dots (middle) lead to  different orientations of the RNA strands. Such difference will add a twist (half turn) at one edge and release a twist from another adjacent edge. Hence it can be seen as a small shift of the strand along adjacent vertices.}
\label{Ajunction}
\end{figure}
The different orientations in the two junction types result from a difference of only a few base pairs in the
 lengths of the RNA strands. Experiments show that the strands are base paired in the middle of the 
 edges of the RNA cage over a length of about 13- 17 nucleotides, but they provide no evidence for 
 basepairing close to the juctions, and any further detail about the structure close to the junctions is 
 unknown \cite{Tang}. 
However if base pairing does not occur close to the junctions 
 the single stranded portions of the RNA allow flexibility for the strands to turn and change into the 
configuration in Fig. \ref{Fig2}(b). The latter was obtained by budges of only 4 nucleotides 
 in the construction of the graph in \cite{JACS}. 
 On the other hand, if there is a  shift in length of  about 5 nucleotides, the new junction type in Fig.~\ref{Fig2} indicates an equally plausible scenario. 
We therefore extend the set of vertex types here and determine the energetically optimal overall 
distribution of vertex configurations in the dodecahedral cage based on this extended set. In comparison to the previous approach, this new way not only leads to a more favourable net result  but also renders twists in addition to those implied by the helical structure  on the edges of the structure  unneccesary. 
\bigskip

As in \cite{Bruinsma}, we assume that all edges of the dodecahedron are occupied by rigid duplex RNA 
segments. Due to the helical nature of the duplex structure, the edge length of the dodecahedral cage 
determines whether an additional twist occurs along each  edge. We therefore demonstrate our 
approach for a particular example, and discuss subsequently how it would have to be modified for the 
case of RNA cages of different sizes. According to results in \cite{Tang} about 1500 of the 4322 
nucleotides of the genome are confined to the dodecahedral cage of pariacoto virus. There are hence 
50 nucleotides located along each of the 30 edges, organised in a double-helical structure of a length of
 25 nucleotides. This implies that there are approximately 2.5 helical turns per edge, resulting in an additional half-turn 
 between the orientations of the RNA at the two respective vertices bordering an edge. 

\begin{figure}[ht]
\begin{center}
\includegraphics[width=6.5cm,keepaspectratio]{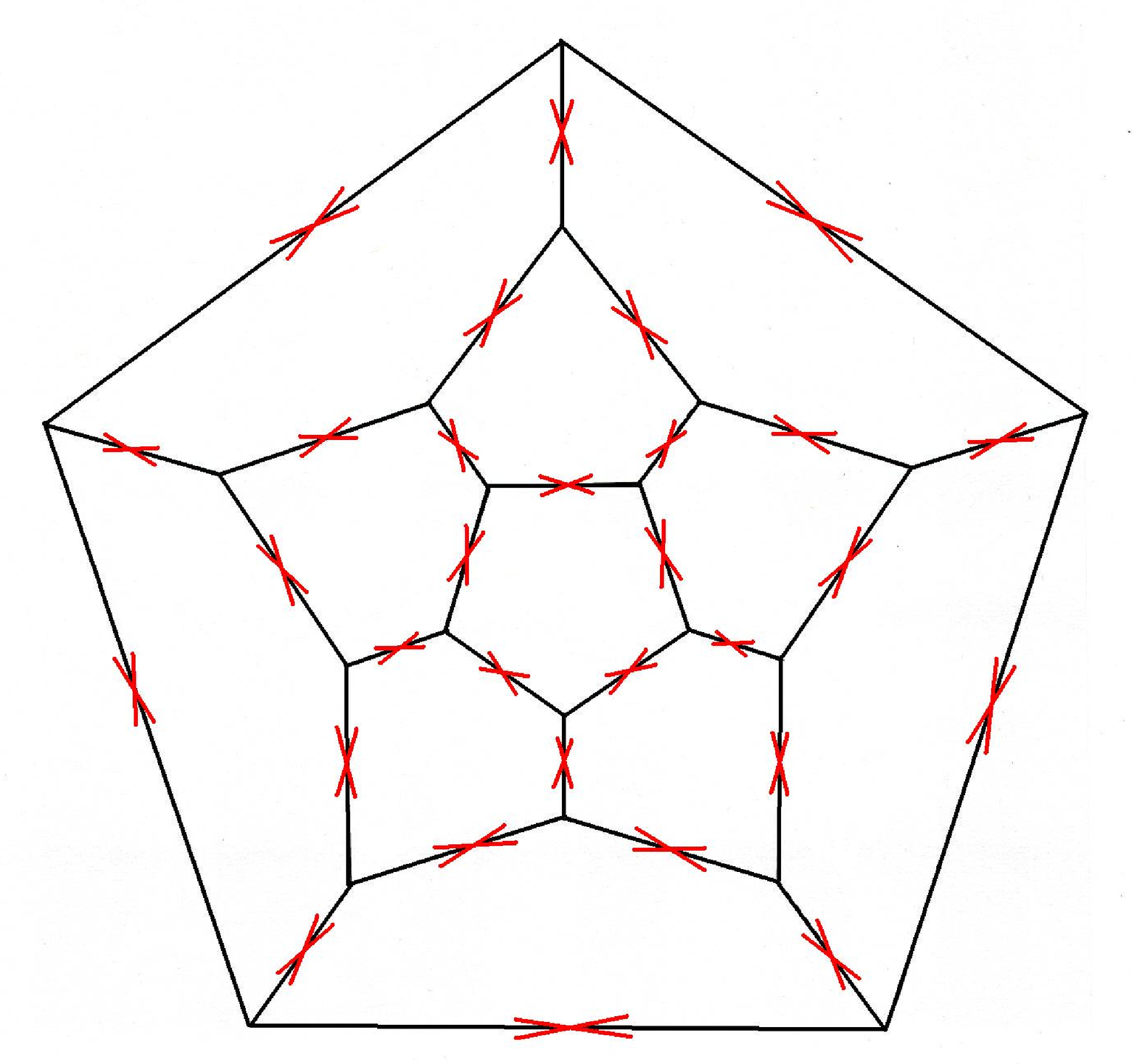}
\end{center}
\caption{A dodecagonal cage in planar projection, with a cross as a schematic representation of the half-turn occurring along this edge.}
\label{Fig3}
\end{figure}

We therefore start our considerations with a dodecagonal cage in planar projection, on which the 
occurrence of this  additional twist along each edge has schematically been indicated by a cross-over 
(see Fig. \ref{Fig3}). 

\begin{figure}[]
\begin{center}
\includegraphics[width=6.5cm,keepaspectratio]{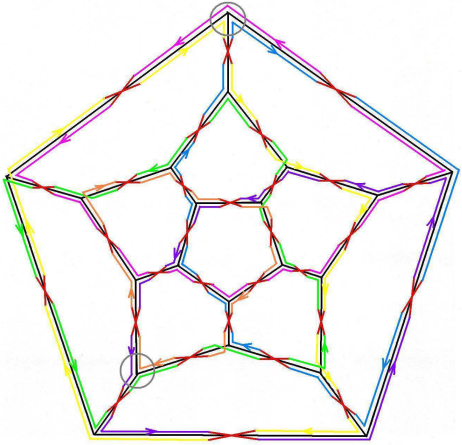}
\end{center}
\caption{The energetically best configuration allowing for separate strands.}
\label{Fig4}
\end{figure}

\begin{figure}[]
\begin{center}
(a) \includegraphics[width=6.5cm,keepaspectratio]{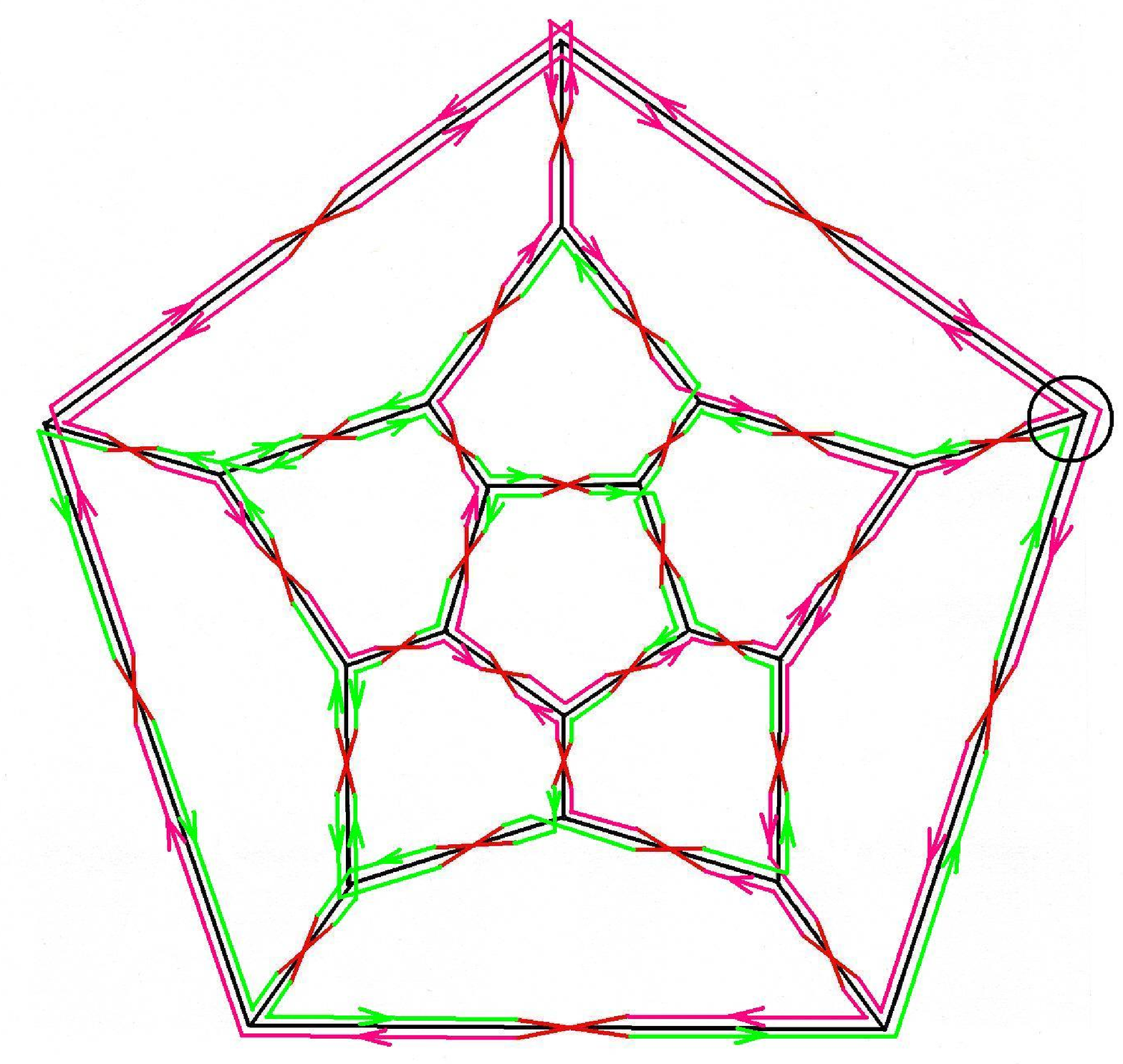} \qquad 
(b) \includegraphics[width=6.5cm,keepaspectratio]{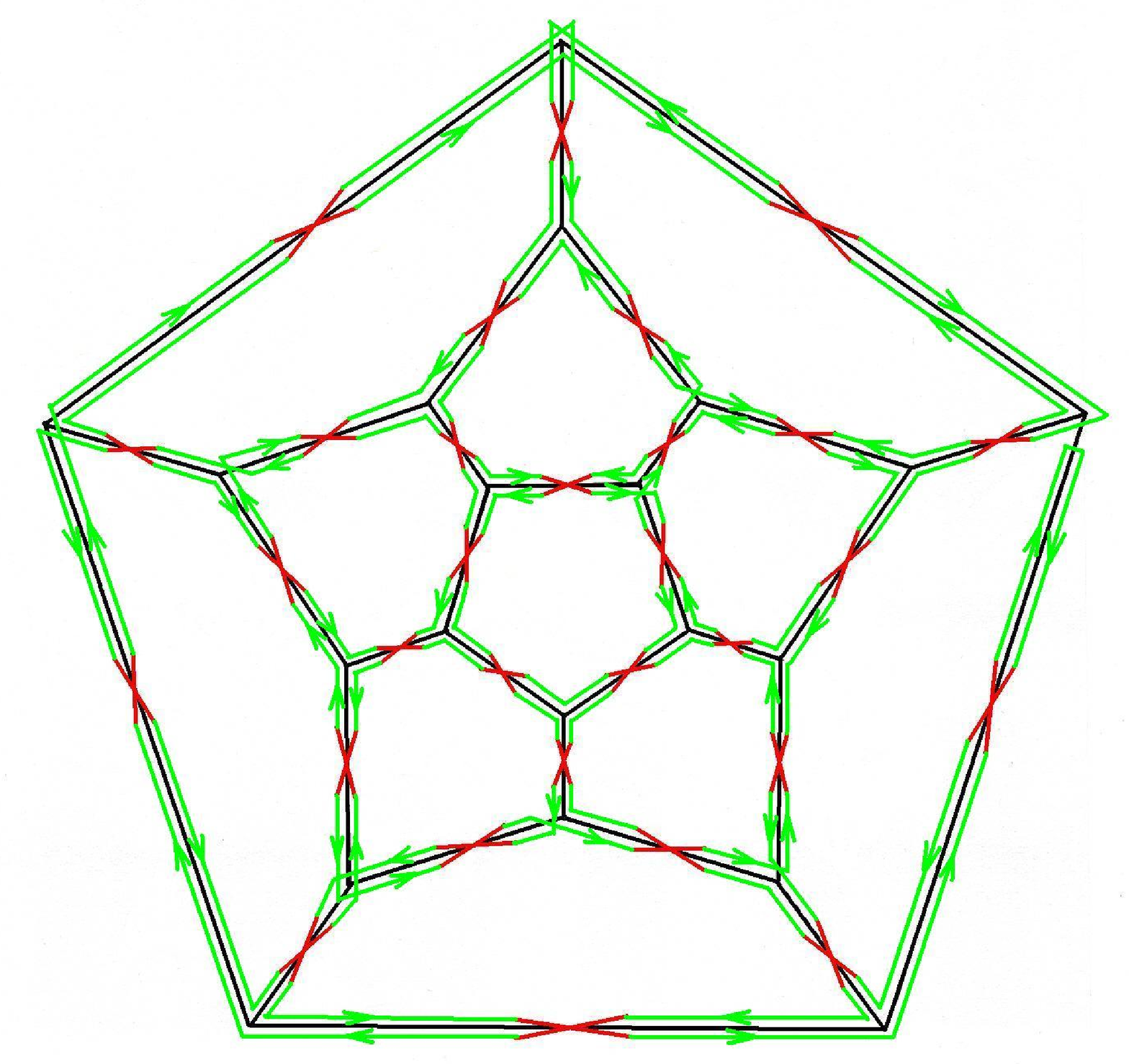}
\end{center}
\caption{The energetically best configuration with two strands in (a), and with one strand in (b).}
\label{Fig5}
\end{figure}

Taking these additional half-turns into account, it is not possible to use only the energetically most 
favourable vertex configuration of type A throughout, even if a different number of independent strands 
were allowed, because it is not possible to have precisely two strands in opposite orientation meeting 
each edge in this case. The energetically best configuration allowing for separate strands corresponds 
to the one shown in Fig.~\ref{Fig4}, which has six separate strands with 14 vertex configurations of type 
A and 6 vertex configurations of the new type in Fig.~\ref{Fig2}(b). Note that the vertices of the new type are necessary as otherwise the twists that appear along the edges lead to inconsistent orientation of the strands.
In order to unite the separate strands, some of the type A vertices have to be replaced by other vertex 
types, and we choose again the new vertex type instead of bubble/hairpins and triple hairpins, because 
it is energetically more favourable than these. Results in \cite{Jonoska} (see Lemma 1) 
provide information on how the 
number of strands changes if a vertex of type A is replaced by the new vertex type. In particular, each 
such replacement unites the three strands that meet at this vertex. Therefore, a minimal number of 
replacements is achieved if two vertices are chosen that have disjoint sets of strands. We therefore 
choose the two vertices that are marked by circles in Fig.~\ref{Fig4}. The resulting configuration is shown 
in  Fig.~\ref{Fig5}(a). 
Since every replacement of a vertex of type A by the new vertex configuration reduces the number of 
strands by two, the number of strands cannot be further reduced in this way. In fact, the formula in \cite{Jonoska} shows that the minimal number of strands to assemble a dodecahedron is 2. 
Therefore, we need tointroduce a different vertex type in order to unite the two strands in Fig.~\ref{Fig5}(a). Since the bubble/hairpins type is the next best option from an energetic point of view, we use it to replace the vertex that is
 marked by a circle in Fig.~\ref{Fig5}(a). We hence obtain the configuration in Fig.~\ref{Fig5}(b),  which,
  by construction, corresponds to the energetically most favourable solution. Since more of the genome 
  is located within the dodecahedral cage, we assume that the bubble in fact extends into the interior of 
  the cage, and hence provides the link between the RNA forming the cage structure and that within. 
\smallskip

\begin{figure}[]
\begin{center}
\includegraphics[width=7cm,keepaspectratio]{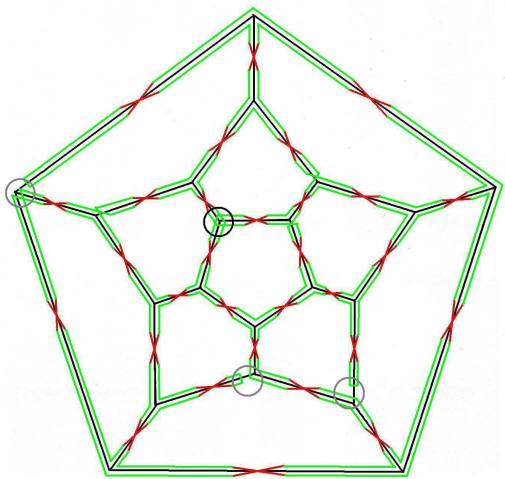}
\end{center}
\caption{An alternative solution involving 3 hairpin/bubble vertices that is of an enegy comparable to that in Fig.~\ref{Fig5}.}
\label{Fig6}
\end{figure}

Many other energetically less favourable configurations are possible. We show one of them in 
Fig.~\ref{Fig6}, which is very close in energy to our optimal solution. 
This structure contains three hairpin/bubbles, 13 vertices of type A and 4 of the new vertices. In 
comparison, it has  hence 2 more vertices of type A and B each, but 4 less of the new vertex type. 
Since the energetics of vertices of type A and those of the new vertex type are similar as we have 
discussed in Fig.~\ref{Ajunction}., its net energy is hence  of a size comparable, but slightly higher, to that
 of our optimal solution in Fig.~\ref{Fig5}.
\bigskip

Our considerations above apply to all viruses that have a dodecahedral cage of a size that 
accommodates an odd number of half turns per edge. Considering that about 5 nucleotides on a single strand
make up one half turn, these
include all viruses with double stranded dodecahedral cages of $300(2k+1)$ nucleotides.
The case of pariacoto virus, that we have used to demonstrate our model, corresponds to the case 
$k=2$. 
\medskip

\begin{figure}[]
\begin{center}
(a) \includegraphics[width=6.5cm,keepaspectratio]{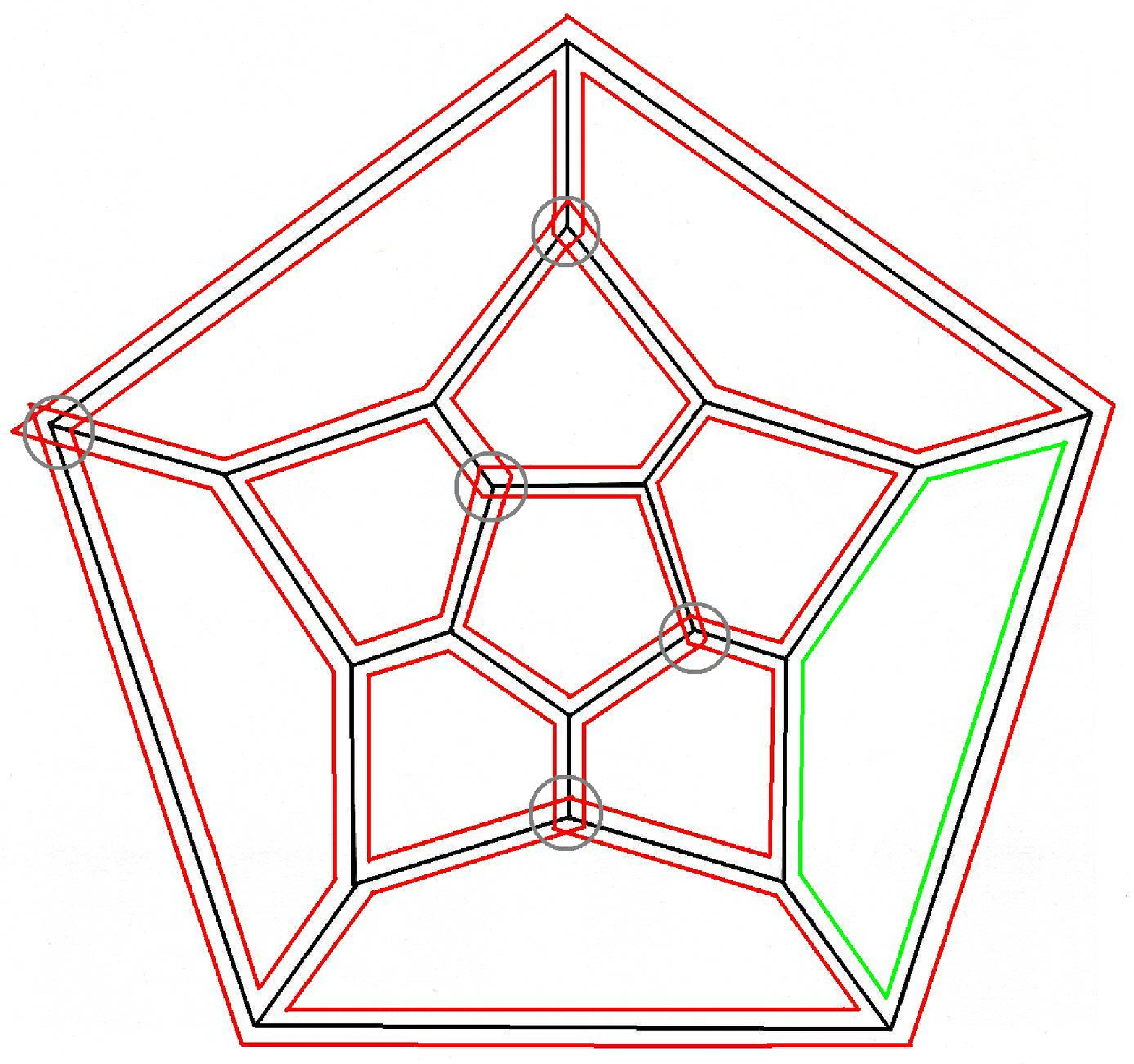} \qquad
(b) \includegraphics[width=6.5cm,keepaspectratio]{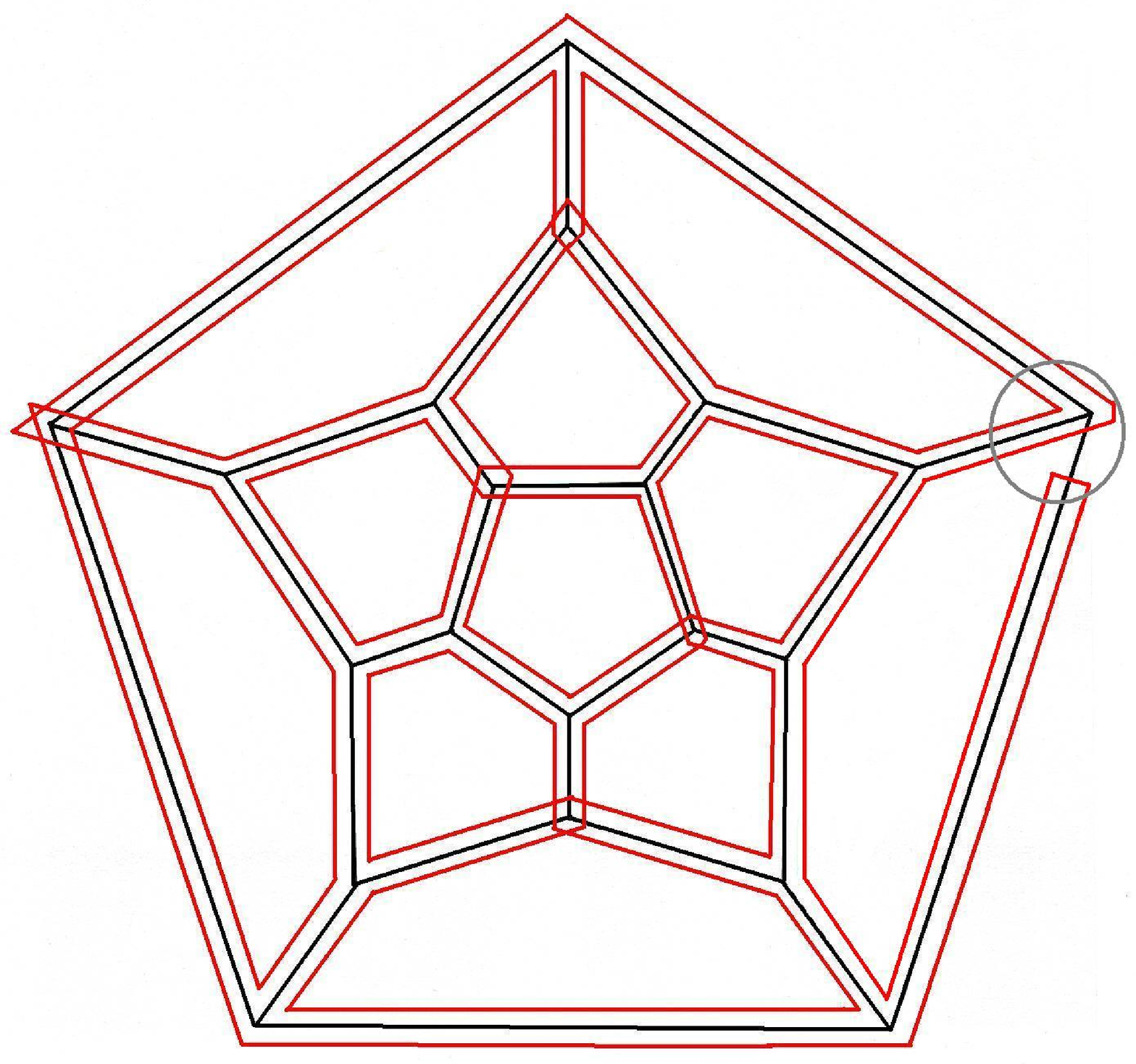}
\end{center}
\caption{A reduction to a two-strand configuration via a replacement of type A vertices by the new vertex type (a), and the final solution after introduction of a bubble/hairpin vertex in (b).}
\label{Fig7}
\end{figure}

We finally discuss the scenario for double stranded dodecahedral   cages with an even number of half turns per edge. 
In these cases, there are no additional twists on the edges, because every edge corresponds to a double-stranded segment of $10k$ nucleotides. As before, we start with a planar representation of the dodecahedron with 
individual strands making the boundaries of each face, preserving opposite orientation to their counterparts. 
 In this case all vertex configurations are of type A. This arrangement corresponds 
 to a model with 12 separate strands which are the maximal number of strands needed to assemble a dodecahedron  \cite{Jonoska}. Results in \cite{Jonoska} imply that $3+2n$, $n\in\N$, loops can be connected 
by a single strand via a replacement of $n+1$, $n\in\N$, vertices of type A by the new vertex type. Hence a replacement of 5 vertices leads to two separate strands as demonstrated in Fig.~\ref{Fig7}(a). The vertices that have been 
replaced are marked by circles, and the two resulting strands are shown in different colours.  

We remark that the exact choice of the vertices does not matter for the final result due to symmetry: one 
always obtains a separate loop in one face, and a large loop connecting all other faces. In order to unite
 these two loops, it is again necessary to introduce a bubble/hairpin vertex configuration. The 
 corresponding solution is shown in Fig.~\ref{Fig7}(b), with the location of the hairpin/bubble indicated by 
 a circle. This configuration corresponds by construction to the energetically optimal solution for viruses 
 with cages of a size of 
300(2k), $k\in\N$, nucleotides. It is energetically more favourable that 
 previous solutions \cite{Bruinsma}, and makes additional twists along the edges unneccesary.

For all cage structures not falling in either of the two families discussed above,
i.e., with edges not being exact multiples of half turns,  additional twists must 
occur on all edges in order to compensate for the relative orientations of the polypeptide chains at the 
two respective boudaries of each edge caused by its helical structure. Such additional twists incur extra 
energy costs and we therefore conjecture that these are less likely to occur in nature. 

We finally comment on the efficiency of extraction of a linear strand from the RNA cage structure. 
We conjecture that the hairpin that appears at the bubble/hairpin vertex is in fact a single stranded portion which extends in the interior of the cage and thus contains free ends. Even though it is very likely that the RNA is knotted, the fact that there is only a very limited amount of base pairing at the edges (maybe just a single (half) twist as only the middle five nucleotides can be confirmed to be base paired \cite{Tang})
the single stranded ends can start untangling from one of the edges and subsequently unwind the whole 
structure. Moreover, it has been proven that topoisomerases such as {\it E. Coli} topo III can knot and unknot single stranded RNA \cite{RNAtopo}. In fact, the activity of topo III is increased at higher temperature when the double helix and supercoiling are relaxed \cite{Wilson}. In the case of the RNA viral cage, if the base pairing at the
vertices is missing, such relaxed and partially denatured substrate is already present and 
the enzyme could perform under optimal conditions.

\section*{Acknowledgements}
RT has been supported by an EPSRC Advanced Research Fellowship. She is very grateful to the 
University of South Florida for financial support of her visit during which the results of this paper have 
been developed. The research of NJ has been partially supported by NSF grants 
 CCF \# 0432009 and CCF \# 0523928.

\end{document}